\documentclass[preprint]{vgtc}            

\onlineid{1129}

\vgtccategory{Research}

\vgtcpapertype{evaluation}

\title{Harnessing LLMs Without Surrendering Control: Delegation Boundaries in Visual Data Storytelling Authoring}

\author{%
  Zhuojun Jiang\thanks{e-mail: zjian115@asu.edu}\\ %
  \scriptsize Arizona State University %
\and
  Yuki Ueno\thanks{e-mail: yueno@asu.edu}\\ %
  \scriptsize Arizona State University %
\and
  Chris Bryan\thanks{e-mail: cbryan16@asu.edu}\\ %
  \scriptsize Arizona State University %
}

\authorfooter{
  \item
  	Zhuojun Jiang, Yuki Ueno and Chris Bryan are with Arizona State University.
  	E-mail: \{zjian115\,$|$\,yueno\,$|$\,cbryan16\}@asu.edu\,.
}

\abstract{
Despite the emergence of large language models (LLMs) for visual data storytelling workflows, there are open questions about how authors decide what activities or tasks to entrust to them and what should be ``protected'' or maintained under human control. To investigate this, we interviewed a cohort of 12 expert visual data storytellers. Our analysis shows that participants rarely treated LLMs as autonomous storytellers. Instead, they tend to selectively delegate execution-oriented tasks to LLMs while retaining control over activities that shape narrative intent and story meaning.  Our findings show that LLM assistance is most productive after human seeding and constraint-setting, and that it shifts labor from production to verification. We discuss design implications for boundary-aware authoring tools, data-grounded generation, low-fidelity ideation, and reporting practices for LLM-based visualization research. Supplemental materials for this paper are available at \url{https://osf.io/hcnp6}.
}


\teaser{
  \vspace{-15pt}
  \centering
  \includegraphics[width=0.8\linewidth]{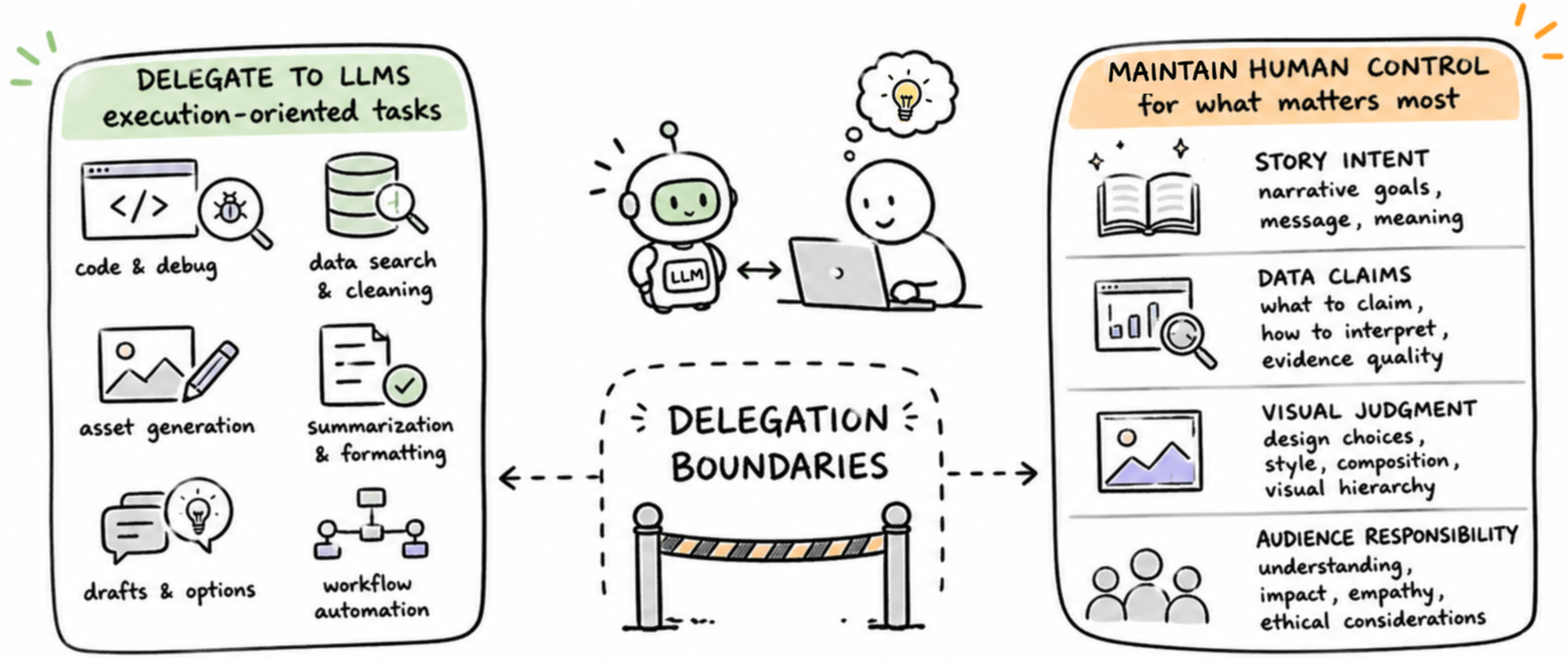}
  \caption{%
  Visual data storytellers harness LLMs by delegating execution-oriented tasks while protecting story intent, data claims, visual judgment, and audience responsibility. We characterize these practices as \textit{delegation boundaries}.%
  }
  \label{fig:teaser}
}

\graphicspath{{figs/}{figures/}{pictures/}{images/}{./}} 

\usepackage{tabu}                      
\usepackage{booktabs}                  
\usepackage{lipsum}                    
\usepackage{mwe}                       

\usepackage{mathptmx}                  

\usepackage{soul}
\usepackage{color}
\usepackage{romannum}

\usepackage{xcolor}
\usepackage{tcolorbox}

\usepackage{enumitem}

\tcbuselibrary{skins}

\definecolor{colUC}{HTML}{93C7F9}
\definecolor{colDL}{HTML}{84B8C4}
\definecolor{colBR}{HTML}{C5E3EB}
\definecolor{colPR}{HTML}{EABBC5}
\definecolor{colVS}{HTML}{D7D3FD}
\definecolor{colAC}{HTML}{F2D78E}

\newcommand{\kwtag}[2]{%
  \tcbox[on line, arc=3pt, outer arc=3pt,
    colback=#1, colframe=#1,
    boxsep=1pt, left=1pt, right=1pt, top=1pt, bottom=1pt,
    boxrule=0pt]{\small\textbf{#2}}%
}

\newcommand{\tagUC}{\kwtag{colUC!60}{UC}}
\newcommand{\tagDL}{\kwtag{colDL!60}{DL}}
\newcommand{\tagBR}{\kwtag{colBR!60}{BR}}
\newcommand{\tagPR}{\kwtag{colPR!60}{PR}}
\newcommand{\tagVS}{\kwtag{colVS!60}{VS}}
\newcommand{\tagAC}{\kwtag{colAC!60}{AC}}

\newcommand{\tagUCx}[1]{\kwtag{colUC!60}{UC\textsubscript{#1}}}
\newcommand{\tagDLx}[1]{\kwtag{colDL!60}{DL\textsubscript{#1}}}

\newcommand{\tagPRx}[1]{\kwtag{colPR!60}{PR\textsubscript{#1}}}
\newcommand{\tagVSx}[1]{\kwtag{colVS!60}{VS\textsubscript{#1}}}
\newcommand{\tagACx}[1]{\kwtag{colAC!60}{AC\textsubscript{#1}}}

\usepackage{tabularx}
\usepackage{array}

\usepackage{amssymb}

\usepackage{tikz}
\newcommand{\dtag}[1]{%
  \begin{tikzpicture}[baseline=(char.base)]
    \node[
      shape=rectangle, 
      fill=gray!20,        
      rounded corners=3pt,
      inner ysep=1.5pt,     
      inner xsep=3pt       
    ] (char) {\footnotesize\textbf{#1}};
  \end{tikzpicture}%
}

\newcommand*{\zj}{\textcolor{black}}

\begin{document}
\pagenumbering{arabic}



\maketitle
\section{Introduction}
\label{sec:intro} 
LLMs and other forms of Generative AI are increasingly being woven into the practices of visualization researchers, designers, and authors~\cite{hutchinson2025foundation, wang2025visualization}. 
In particular, visual data storytelling has emerged as a potential avenue for this type of work~\cite{he2025leveraging}. Prior work investigating the potential of human-GenAI collaboration for visual data storytellers has identified many potential activities and benefits.
For example, Li et al.~\cite{li2025ai} interviewed 18 data workers (primarily researchers and business analysts) to explore their preferences for potential human-AI collaboration during storytelling planning, implementation, and communication, such as generating candidate story ideas and plot points, 
sourcing and summarizing background material and datasets, 
and producing and debugging code, and categorizing the potential roles an LLM can occupy during such collaboration (e.g., as a creator, optimizer, reviewer, or assistant).

However, gaps remain for understanding current expert practice.
For one, the authoring of visual data stories depends on a chain of propagating decisions, such as first ideating the base story idea and plot points, identifying and collecting data to support the story's intent and messaging, designing and constructing charts, narrative elements, and other assets, and framing the story content to be informative, engaging, and impactful to the intended audience\cite{stolper2016emerging}.
In the abstract, LLMs can potentially assist (or altogether automate) these types of activities, but there are also inherent risks: a model might make ``incorrect'' decisions (i.e., going against the author's wishes), it might overstate claims or hallucinate information or sources, and it might homogenize style or misconstrue the story's intents or goals, and more.
Prior work has provided little nuance into the ``delegation boundaries'' of when and where an author might entrust the LLM enough to surrender control vs. activities where an author would want to maintain ``ownership'' and control.

Along these lines, given the rapid evolution, maturation, and integration of LLMs and Generative AI into daily life, many data storytelling authors (who are already computationally literate) have high familiarity and comfort with LLMs, and regularly use them in their day-to-day activities. For example, the study in~\cite{li2025ai} was conducted in Q1 2023 where only some participants had in depth knowledge and regular usage of AI. Today, LLM usage is widespread, both among academics (over 80\% of researchers reported using LLMs in their workflows in a 2024 survey~\cite{liao2024llms}) and with the general public (a July 2025 survey by AP-NORC found over 70\% of US adults under 30 use AI at least some of the time for information-driven tasks~\cite{apnorc}). As such, the needs, concerns, and attitudes of storytelling authors might likewise be shifting.

At a high level, we seek to understand where and how current visual data storytellers harness the capabilities of LLMs while still maintaining creative and design control\zj{, a set of practices we characterize as delegation boundaries (\cref{fig:teaser}).}
We structure this around a pair of related research questions: 
(\Romannum{1}) \textit{How do current ``LLM proficient'' visual data storytellers incorporate LLMs into their authoring practices?} (\Romannum{2}) \textit{What tasks, decisions, and responsibilities do they delegate, constrain, verify, or protect?} 

To investigate these, we conducted semi-structured interviews with 12 visual storytelling authors with demonstrated experience in creating and publishing public visual data stories. \zj{The collected feedback was analyzed via qualitative coding, pattern analysis, and thematic analysis.} This process allows us to study not only where LLMs are used by actual storytellers, but also understand how and why these authors form delegation boundaries for LLM assistance in the context of their own work (e.g., what can be delegated, what must be verified, what should remain fully human-controlled, etc.).

Broadly, our findings indicate there are a number of \textit{execution-oriented tasks} that authors trust delegating to LLMs, while concepts that focus on narrative intent, conceptualization, and story framing must be protected (i.e., the authors maintain human control over these types of activities). Along these lines, when activities are entrusted to LLMs, author labor and effort primarily shifts from production to verification and risk management, not just for hallucination but also for a number of discrete storytelling and messaging-related risks such as maintaining the appropriate story voice.
Based on our findings, we derive a set of design implications for future work and opportunities about how delegation boundaries and authorial controls can be leveraged to improve data storytelling processes.
\vspace{-14pt}
\section{Related Work}



\zj{Generative AI is reshaping creative work across writing, design, and visual media, prompting growing study of how human and AI agency are negotiated in co-creation\cite{zhang2025exploring,peng2026designtrace}. Visual data storytelling falls within this broader creative practice and refers to communicative experiences that integrate data, visualizations, and narrative elements~\cite{lee2015more}.} In their seminal paper, Segel and Heer~\cite{segel2010narrative} outlined a design space for narrative visualizations, identifying several salient dimensions of visual storytelling (e.g., how chart designs and interactions can enforce structure and narrative flow). Subsequent work, such as Hullman and Diakopoulos~\cite{hullman2011visualization}, examines how rhetorical techniques applied to storytelling visualizations can shape the interpretations and messaging received by readers.

Today, storytelling remains an active area within the visualization community, spanning the development of new tools to produce data stories~\cite{chen2024how,li2024we,shi2020calliope} and studies of their impacts on readers~\cite{shao2024data}. There also exist several peer-reviewed contests and venues for visual data storytelling, including the PacificVis Data Storytelling Contest\footnote{\url{https://visstory.github.io/}} and the Information is Beautiful Awards\footnote{\url{https://www.informationisbeautifulawards.com/}}.

With the emergence of LLMs, the data storytelling community is  now exploring the potential of human-AI collaboration~\cite{he2025leveraging}. In addition to the study~\cite{li2025ai} mentioned in \cref{sec:intro}, several recent efforts have developed human-AI collaborative tools for creating or refining visual data stories (e.g.,~\cite{li2025reflection} surveys 27 recent tools to map the types of patterns and activities currently supported). However, as noted in \cref{sec:intro}, open questions remain about when and where authors actually want to collaborate with LLMs or delegate authority to them, which the current study is designed to address.
\vspace{-4pt}
    
\section{Methods}
\textbf{Participants:} 
We recruited 12 data storytelling authors through criterion-based purposive sampling. Eligible participants had either published work in visual data storytelling (in IEEE TVCG or ACM SIGCHI) or had been shortlisted or awarded in the PacificVis Visual Data Storytelling Contest. Several participants held overlapping identities as researchers, designers, and data story authors (see \cref{tab:participants}), fulfilling both criteria. This sampling strategy was intended to recruit experienced public-facing authors rather than internal data workers (e.g., those working within an organization) or casual visualization creators. 

\begin{table}[t]
    \centering
    \small
    \caption{Participant demographics.}
    \vspace{-9pt}
    \setlength{\tabcolsep}{3pt}
    \begin{tabular}{l l p{2.6cm} c l c}
    \toprule
    ID & Region & Role & Exp. (yrs) & LLM Usage & PVIS\\
    \midrule
    P1  & U.S. & Both (research \& practice) & 5+ & Daily & \checkmark \\
    P2  & U.S. & Academic researcher & 5+ & Daily & \checkmark \\
    P3  & U.S. & Industry practitioner & 3-5 & Occasionally & \checkmark \\
    P4  & U.S. & Both (research \& practice) & 3-5 & Weekly & \checkmark \\
    P5  & U.S. & Industry practitioner & 3-5 & Daily & \checkmark \\
    P6  & Asia & Both (research \& practice) & <3 & Daily & \checkmark \\
    P7  & Asia & Adjacent field & <3 & Daily & \checkmark \\
    P8  & Asia & Academic researcher & 5+ & Daily &  \\
    P9  & Asia & Academic researcher & 5+ & Daily &  \\
    P10 & Asia & Industry practitioner & 5+ & Daily & \checkmark \\
    P11 & Asia & Academic researcher & 3-5 & Weekly &  \\
    P12 & Asia & Industry practitioner & <3 & Daily & \checkmark \\
    \bottomrule
    \end{tabular}
    \vspace{-4pt}
    \begin{minipage}{0.9\linewidth}
    {Asia includes Mainland China, Hong Kong, South Korea, and Singapore.\\
    \checkmark: PacificVis visual data storytelling contest award or shortlisted participant}
    \end{minipage}    
    \label{tab:participants}
    \vspace{-15pt}
\end{table}




\textbf{Procedure:}
\zj{We conducted 20--35 minute semi-structured interviews via online conferencing platforms (e.g., Zoom), focusing on recent works rather than exhaustive career histories. Participants described their narrative visualization or data storytelling experience, authoring workflows, and their use of LLMs. We grounded discussions in concrete works, including ideation, data sourcing, story flow, visual mapping, text writing, coding, asset generation, verification, and final publication. Follow-up questions probed why participants used or avoided LLMs at specific stages, how they evaluated outputs, and how LLM use shaped their sense of authorship, responsibility, and creative control. This study was approved by ASU’s IRB.}

\begin{figure*}[h]
  \centering
  \includegraphics[width=0.95\linewidth]{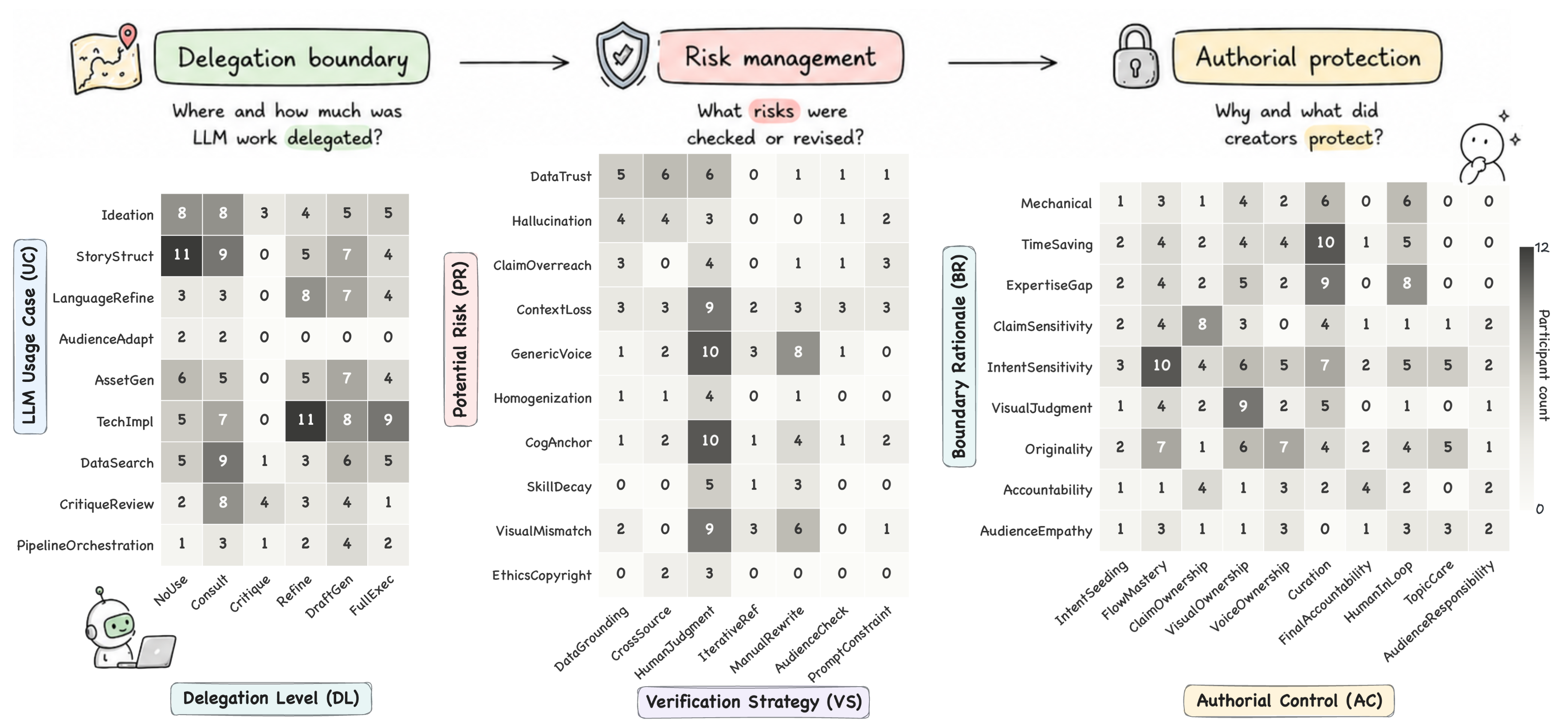}
  \vspace{-5pt}
  \caption{\textbf{From coding framework to pattern analysis.} We organized our codebook's six dimensions into three co-occurrence matrices: delegation boundary (\tagUC$\times$\tagDL), risk management (\tagPR$\times$\tagVS), and authorial protection (\tagBR$\times$\tagAC). Each cell indicates how many of the 12 participants discussed a given code pair at least once. Codes are non-exclusive. \zj{See the supplemental materials for the full codebook.}}
  \label{fig:matrix-overview}
  \vspace{-17pt}
\end{figure*}

\textbf{Coding Process:}
We transcribed and analyzed all interviews\cite{adams2015conducting} with an iterative qualitative coding process. We first segmented the transcripts, then divided longer turns into meaning units when participants discussed multiple ideas in the same response. In the first cycle, we coded meaning units along six dimensions: 
\vspace{-2pt}
\begin{itemize}[noitemsep, leftmargin=1.2em, rightmargin=1em]
    \item \textbf{LLM use case}~(\tagUC): identifies where LLMs entered a storytelling workflow
    \item \textbf{Delegation level}~(\tagDL): captures how much authority participants gave to LLMs during a specific task
    \item \textbf{Perceived risk}~(\tagPR): captures concerns that arose when LLMs were used or considered
    \item \textbf{Verification strategy}~(\tagVS): identifies how participants checked, constrained, revised, or rejected LLM outputs
    \item \textbf{Boundary rationale}~(\tagBR): explains why participants delegated or protected a task
    \item \textbf{Authorial control}~(\tagAC): captures how participants maintained authorship and responsibility when using LLMs
\end{itemize}

Codes in these dimensions capture both where and how participants used LLMs in their workflows, as well as the strategic considerations that shaped those decisions (such as where to delegate authority, how to manage risk, and how to maintain control over key story elements). We refined the codebook iteratively through team discussion until recurring patterns were consistently captured. The final codebook contains the six dimensions and 47 individual codes, and is provided in the supplemental materials.

\textbf{Pattern Analysis Process:}
In the second cycle, we conducted axial coding and cross-case comparison to identify relationships across dimensions and codes. Specifically, we organized our six dimensions into three co-occurrence matrices (see \cref{fig:matrix-overview}), where each matrix row and column represents the codes for that corresponding dimension. \zj{Each matrix pairs a ``where/what'' dimension with its corresponding ``how/why'' dimension to form an analytically coherent question.} We used these matrices to formatively identify patterns across participant responses, briefly summarized below:

(i)~In the \textit{delegation boundary} matrix, LLM use cases that were more execution-oriented, such as generating the technical code to implement a visualization (TechImpl) or generating story assets such as illustrations or UI elements (AssetGen), were highly associated with the refinement, drafting, and full execution delegation levels (i.e., participants entrusted delegating these activities to an LLM). In contrast, participants tended to ``protect'' or proactively maintain control over activities more focused on shaping the narrative intents, concepts, and overall story meaning, such as organizing the story flow, narrative logic, or chapter sequence (StoryStruct), critiquing and reviewing the story (CritiqueReview), or even first originating the ideas or themes for the story (Ideation). Here, the LLM was often either not used or relegated to a consulting role.

(ii)~In the \textit{risk management} matrix, risks such as data trust, hallucination, claim overreach (e.g., an LLM suggesting conclusions beyond what the data supports), and context loss (misunderstanding the story background, audience, or purpose) were often paired with human-driven verification strategies such as source checking, data grounding, adding prompt constraints, and human judgment. Likewise, risks relating to messaging and communicative framing such as generic voice, cognitive anchoring, and visual mismatches (i.e., generating visualizations that failed to match expectations or desired designs) were similarly seen as requiring human oversight and manual editing.

(iii)~In the \textit{authorial protection} matrix, authors emphasized maintaining control over high-level narrative concept and intent, such as how the core storyline flows (FlowMastery), what claims and conclusions are made (ClaimOwnership), what the visual designs, compositions, and mappings should ultimately look like (VisualOwnership), and, ultimately, what the final product should look like (FinalAccountability). These were often linked to rationales such as that the task required the author's intent, perspective, and stance (IntentSensitivity), that some claims required human validation (ClaimSensitivity), that LLM outputs might become overly homogenized or clich\'{e}d (Originality), and that final responsibility in the story lay with the author (Accountability).
\vspace{-4pt}

\section{Thematic Analysis Results}

Based on the co-occurrence analysis, we subsequently conducted a thematic analysis of relevant participant comments that dealt with emergent patterns. We discuss identified themes below.

\textbf{Execution Is Delegated, Narrative Is Protected.}
Participants consistently treated execution-related activities, such as coding, debugging, data processing, and routine technical implementations (e.g., displaying a tooltip), as delegable work. In contrast, they wanted control over the conceptualization and narrative framing of the story, protecting aspects such as how it should flow, when and how visuals should be mapped into the story, how data should be interpreted, and overall design direction. 

Several quotes reflected these sentiments. P1 described a selective delegation strategy: \textit{``For deciding the theme, I would ask the LLM to list potential ideas. But in the design phase, I would still discuss with my partner and draw the abstract sketch on paper''} (\tagUCx{Ideation}$\times$\tagDLx{Consult}, \tagUCx{StoryStruct}$\times$\tagDLx{NoUse}).
Similarly, P6 and P7 adopted a "design-first, delegate-later" approach. For P6, this made up for their coding weaknesses,
P7 emphasized that they and the other story creators already knew which charts were appropriate (i.e., in terms of visualization design principles and selection of marks and channels), but that \textit{``the code part was all written with the LLM''}  (\tagUCx{TechImpl}$\times$\tagDLx{FullExec}).
Some authors, like P2, were open to AI-assisted narrative drafting (\tagUCx{StoryStruct}$\times$\tagDLx{DraftGen}), \zj{but \textit{``for design, I probably still would not ask it''} (\tagUCx{AssetGen}$\times$\tagDLx{NoUse}).} P9 and P12 mentioned they could have the LLM suggest feedback on \textit{aspects} of story ideas (\tagUCx{Ideation}$\times$\tagDLx{Critique}), but such decisions were ultimately maintained and originated by themselves.
In this sense, the idea of delegation boundaries can be understood as degrees of authority: LLMs may propose, draft, or execute, but creators retain authority over final narrative and visual decisions.

\textbf{LLMs Work Best After a Human Seed or Constraint.}
Participants rarely described LLMs as useful blank-slate storytellers. Instead, they emphasized the need to seed or constrain the model with a topic, standpoint, draft, design system, or audience specification before generation. This practice ensures LLM assistance remains predictable and directed, preserving authorial responsibility for the narrative core. 
P6 prepared \textit{``the narrative outline and the design specifications before communicating with the LLM,''} while P12 would use AI in early-stage research but not hand over the \textit{``main conception,''} noting that the model worked better when given a draft to optimize (\tagACx{IntentSeeding}, \tagACx{FlowMastery}). They adopted a "constraint-first" workflow and P11 noted that raw data alone rarely produces desired results, but a clear perspective in the prompt does (\tagACx{ClaimOwnership}).
Prompting therefore became a design activity, with participants configured the space in which the model was allowed to operate. This pre-generation control made things more predictable, but it also shaped the range of ideas the model could return. P10 explained: \textit{``It expanded some of my thinking, especially the logic between events... But it is still based on the angle I give it, and it tends to go along with me... It's different from discussing with a person''} (\tagPRx{CogAnchor}). Thus, prompt constraints cut both ways.

\textbf{LLMs Shift Effort from Production to Verification.}
LLM assistance did not straightforwardly reduce effort while it shifted work from producing artifacts to checking, constraining, and correcting outputs. 
For data sourcing, participants emphasized verification over direct acceptance. 
P1 explained: \textit{``It's not good at finding good data sources online... I need to check the accuracy or trustworthiness of the data''} (\tagPRx{DataTrust}$\times$\tagVSx{DataGrounding}).
P6 applied two tests: whether a source link was real, and whether its conclusion fit (\tagPRx{Hallucination}$\times$\tagVSx{CrossSource}).
Verification extended beyond factual checking to tone, narrative scope, and the data-story relationship. 
P4 described iterative revision because the model \textit{``changes the tone''} from what was intended (\tagPRx{ContextLoss}$\times$\tagVSx{IterativeRef}, \tagVSx{HumanJudgment}). 
P10 captured the role shift: \textit{``Now I become someone who reviews its content output, rather than someone who revises my own work... I lose part of the joy and part of the ability to train my own thinking''} (\tagPRx{SkillDecay}$\times$\tagVSx{HumanJudgment}, \tagVSx{ManualRewrite}). 


\textbf{Authorial Control Is Maintained Through Curation, Not Total Manual Creation.}
Participants did not define authorship as making every component manually. Instead, they maintained authorial control by selecting, constraining, revising, rejecting, and integrating model outputs. This distinction matters because it avoids a simplistic division between \textit{human-made} and \textit{delegated} data stories. P3, who used multiple tools for text, images, and interface generation, said: \textit{``I would not consider them as the final outputs. I still need to add a human touch''} (\tagACx{Curation}). P10 interestingly described the final story as a ``collision'' between the human creator and the model: \textit{``AI is one variable, and I am another. What comes out is produced by the collision between my background and thinking, and its knowledge base.''} 
P6 identified narrative ordering, transitions, and causal structure as
irreducibly requiring human judgment: \textit{``The arrangement of the framework, the transitional sentences, and the cause-and-effect relationship between parts, these definitely need a person to judge... It is a kind of director's thinking, an embodied experience''} (\tagACx{FlowMastery}, \tagACx{ClaimOwnership}, \tagACx{HumanInLoop}). This curatorial view of authorship also appeared in P10's claim that a creator's \textit{``attention and care for the topic''} and \textit{``initiative and attitude''} are difficult to be replaced (\tagACx{VoiceOwnership}, \tagACx{HumanInLoop}, \tagACx{TopicCare}).

\textbf{Risks Go Beyond Hallucination.}
All participants cited hallucination as a concern, but other risks were equally
salient, such as cognitive anchoring, premature convergence, style homogenization, and unclear provenance. 
P2 worried that fluent-sounding output could mask weak reasoning: \textit{``I can be led along by it... Sometimes what it says looks very logical, but when I read it again, I find that it isn't really logical''} (\tagPRx{Hallucination}). 
P5 described a related risk of premature convergence, where the model \textit{``quickly converges the idea''} and may \textit{``interrupt my inspiration''} (\tagPRx{CogAnchor}), while
P12 connected this to homogenization and being trapped inside the model's frame,
noting difficulty \textit{``jumping out of the current frame''} (\tagPRx{Homogenization}).
P9 tied extended use to skill decay: \textit{``It can help you think of more
things, but it can also limit you... If you use it too much, you become dumber
yourself''} (\tagPRx{SkillDecay}).
These risks matter for data storytelling specifically because story quality
depends not only on factual correctness, but on authentic audience connection, 
which relies on trust in the story's evidence, framing, and intent.
\vspace{-4pt}

\section{Discussion and Conclusion}
Based on our analysis, we briefly summarize a set of \textit{design implications} for creators, researchers, and tool builders navigating the shift toward LLM-assisted visual data storytelling:

\dtag{I1} \textit{Proactive boundary-aware authoring.} 
Human-LLM storytelling is not a binary choice but often a nuanced negotiation, with delegation levels varying depending on several factors including the specific activities, perceived potential risks, and desired authorial control. 
Authors should proactively identify which story components should remain under human control to mitigate perceived risks, which adopts a position of preserving human agency without rejecting automation~\cite{amershi2019guidelines,heer2019agency}.

\dtag{I2} \textit{Data-grounded generation and provenance-aware verification.} 
Our participants treated model outputs as hypotheses to be checked rather than facts to be accepted, especially when outputs involved claims, causal explanations, or audience-facing content. This reframes verification as a part of authorship: maintaining ownership requires checking whether narrative statements and visualizations remain accurately grounded in data, context, and intent\cite{ragan2015characterizing}. For researchers, this points to a need for methods that support improving verification activities~\cite{fan2026help}; e.g., being able to quantify not only story quality but also \textit{claim-to-data alignment}, such as flagging unsupported claims, linking narrative elements to constructed visualization, and justifying suggestions for narrative and visual elements.


\dtag{I3} \textit{Treat prompt constraints as first-class design artifacts.} Participants often maintained control before generation by specifying outlines, design systems, API scopes, or narrative requirements in their prompts\cite{arawjo2024chainforge}, which included defining audience requirements, tone, data scope, visual style, evidence requirements, and prohibited claims\cite{liu2022design}. Future tools and authoring practices should likewise treat such constraints as persistent design artifacts, as a way to promote a curatorial view of authorship. 

\dtag{I4} \textit{Support tarnish and pushback.} Several participants worried that polished LLM outputs could narrow the creative space too early, producing cognitive anchoring and premature closure~\cite{cho2017anchoring}. This suggests that early-stage LLM collaboration should explicitly not aim for finished prose or publication-worthy visuals. Authors can also engage with models as sparring partners by asking for counter-arguments, alternatives, skeptical audience reactions, or competing visual metaphors. Future work can examine how the fidelity of AI suggestions affects exploration breadth, particularly in early-stage storytelling\cite{sefelin2003paper}. Likewise, future tools can support low-fidelity AI modes that deliberately keep outputs rough, plural, and provisional, as a way to defer closure and promote creativity\cite{wadinambiarachchi2024effects}.

\textbf{Limitations:}
Our study recruitment focused on expert and recognition-based creators with research or storytelling contest experience, which may make them especially attentive to originality, visual polish, and the legitimacy of authorial contribution, so the findings may not fully generalize to other storytelling settings such as casual data storytellers or company-focused data workers. Our study also relies on self-reported interviews. Although participants grounded their responses in concrete workflows, tools, and authoring episodes, future work could 
combine interviews with longitudinal observation or controlled authoring studies. Participants also discussed LLMs together with adjacent generative AI tools, including image generation and AI-assisted interface tools. This reflects real creative workflows, where creators combine multiple AI systems, but it also means that some findings concern LLM-mediated generative AI practice rather than text-only LLM use \cite{tang2026human}. 
Finally, given the rapid evolution of LLMs, these delegation boundaries themselves will dynamically evolve alongside emerging community norms regarding AI usage and disclosure.

\textbf{Conclusion:}
LLMs are becoming part of authoring visual data stories, but authors are reticent to turn to full autonomy. Instead, authors harness LLMs selectively: they tended to delegate execution-heavy tasks, but they also preserved significant human responsibility and oversight for many activities related to the narrative intent, curation, and ideation/creativity.
These \textit{delegation boundaries} can help explain how creators benefit from LLMs without surrendering authorial control. They also reveal new design challenges: tools must support not only generation, but also constraint-setting, provenance-aware verification, transparent reporting, and even pushback and tarnish as a way to promote authenticity. Put another way, LLMs change visual data story authoring not by replacing authorship, but by making authorial control something that must be actively designed, negotiated, and verified.

\bibliographystyle{abbrv-doi-hyperref}

\bibliography{template}

@article{li2025ai,
    author={Li, Haotian and Wang, Yun and Liao, Q. Vera and Qu, Huamin},
    journal={IEEE Transactions on Visualization and Computer Graphics}, 
    title={{Why is AI Not a Panacea for Data Workers? An Interview Study on Human-AI Collaboration in Data Storytelling}}, 
    year={2025},
    volume={31},
    number={10},
    pages={7598-7613},
    doi={10.1109/TVCG.2025.3552017}
}

@ARTICLE{segel2010narrative,
  author={Segel, Edward and Heer, Jeffrey},
  journal={IEEE Transactions on Visualization and Computer Graphics}, 
  title={{Narrative Visualization: Telling Stories with Data}}, 
  year={2010},
  volume={16},
  number={6},
  pages={1139-1148},
  doi={10.1109/TVCG.2010.179}}

@ARTICLE{lee2015more,
  author={Lee, Bongshin and Riche, Nathalie Henry and Isenberg, Petra and Carpendale, Sheelagh},
  journal={IEEE Computer Graphics and Applications}, 
  title={{More Than Telling a Story: Transforming Data into Visually Shared Stories}}, 
  year={2015},
  volume={35},
  number={5},
  pages={84-90},
  doi={10.1109/MCG.2015.99}}

@misc{wang2025visualization,
	title = {Visualization {Generation} with {Large} {Language} {Models}: {An} {Evaluation}},
	shorttitle = {Visualization {Generation} with {Large} {Language} {Models}},
	url = {http://arxiv.org/abs/2401.11255},
	doi = {10.48550/arXiv.2401.11255},
	urldate = {2026-01-05},
	publisher = {arXiv},
	author = {Wang, Xinyu and Liang, Chenwei and Zheng, Shunyuan and Liang, Jinyuan and Li, Guozheng and Zhang, Yu and Liu, Chi Harold},
	month = dec,
	year = {2025},
	note = {arXiv:2401.11255 [cs]},
}

@article{hutchinson2025foundation,
	title = {Foundation model assisted visual analytics: {Opportunities} and {Challenges}},
	volume = {130},
	issn = {0097-8493},
	shorttitle = {Foundation model assisted visual analytics},
	url = {https://www.sciencedirect.com/science/article/pii/S0097849325000871},
	doi = {10.1016/j.cag.2025.104246},
	urldate = {2026-04-01},
	journal = {Computers \& Graphics},
	author = {Hutchinson, Maeve and Jianu, Radu and Slingsby, Aidan and Madhyastha, Pranava},
	month = aug,
	year = {2025},
	pages = {104246},
}

@ARTICLE{he2025leveraging,
  author={He, Yi and Xu, Ke and Cao, Shixiong and Shi, Yang and Chen, Qing and Cao, Nan},
  journal={IEEE Transactions on Visualization and Computer Graphics}, 
  title={Leveraging Foundation Models for Crafting Narrative Visualization: A Survey}, 
  year={2025},
  volume={31},
  number={10},
  pages={9303-9323},
  doi={10.1109/TVCG.2025.3542504}
}

@inproceedings{amershi2019guidelines,
    author = {Amershi, Saleema and Weld, Dan and Vorvoreanu, Mihaela and Fourney, Adam and Nushi, Besmira and Collisson, Penny and Suh, Jina and Iqbal, Shamsi and Bennett, Paul N. and Inkpen, Kori and Teevan, Jaime and Kikin-Gil, Ruth and Horvitz, Eric},
    title = {Guidelines for Human-AI Interaction},
    year = {2019},
    isbn = {9781450359702},
    publisher = {Association for Computing Machinery},
    address = {New York, NY, USA},
    url = {https://doi.org/10.1145/3290605.3300233},
    doi = {10.1145/3290605.3300233},
    booktitle = {Proceedings of the 2019 CHI Conference on Human Factors in Computing Systems},
    pages = {1–13},
    numpages = {13},
    location = {Glasgow, Scotland Uk},
    series = {CHI '19}
}

@article{heer2019agency,
    author = {Jeffrey Heer },
    title = {Agency plus automation: Designing artificial intelligence into interactive systems},
    journal = {Proceedings of the National Academy of Sciences},
    volume = {116},
    number = {6},
    pages = {1844-1850},
    year = {2019},
    doi = {10.1073/pnas.1807184115},
    URL = {https://www.pnas.org/doi/abs/10.1073/pnas.1807184115},
    eprint = {https://www.pnas.org/doi/pdf/10.1073/pnas.1807184115}
}

@ARTICLE{chen2024how,
  author={Chen, Qing and Cao, Shixiong and Wang, Jiazhe and Cao, Nan},
  journal={IEEE Transactions on Visualization and Computer Graphics}, 
  title={How Does Automation Shape the Process of Narrative Visualization: A Survey of Tools}, 
  year={2024},
  volume={30},
  number={8},
  pages={4429-4448},
  doi={10.1109/TVCG.2023.3261320}}

@inproceedings{shao2024data,
author = {Shao, Hongbo and Martinez-Maldonado, Roberto and Echeverria, Vanessa and Yan, Lixiang and Gasevic, Dragan},
title = {Data Storytelling in Data Visualisation: Does it Enhance the Efficiency and Effectiveness of Information Retrieval and Insights Comprehension?},
year = {2024},
isbn = {9798400703300},
publisher = {Association for Computing Machinery},
address = {New York, NY, USA},
url = {https://doi.org/10.1145/3613904.3643022},
doi = {10.1145/3613904.3643022},
booktitle = {Proceedings of the 2024 CHI Conference on Human Factors in Computing Systems},
articleno = {195},
numpages = {21},
location = {Honolulu, HI, USA},
series = {CHI '24}
}

@inproceedings{fan2026help,
    author = {Fan, Guangrui and Liu, Dandan and Pan, Lihu and Zhang, Rui},
    title = {When Help Hurts: Verification Load and Fatigue with AI Coding Assistants},
    year = {2026},
    isbn = {9798400722783},
    publisher = {Association for Computing Machinery},
    address = {New York, NY, USA},
    url = {https://doi.org/10.1145/3772318.3791176},
    doi = {10.1145/3772318.3791176},
    articleno = {733},
    numpages = {25},
    location = {
    },
    series = {CHI '26}
}

@inproceedings{liu2022design,
author = {Liu, Vivian and Chilton, Lydia B},
title = {Design Guidelines for Prompt Engineering Text-to-Image Generative Models},
year = {2022},
isbn = {9781450391573},
publisher = {Association for Computing Machinery},
address = {New York, NY, USA},
url = {https://doi.org/10.1145/3491102.3501825},
doi = {10.1145/3491102.3501825},
booktitle = {Proceedings of the 2022 CHI Conference on Human Factors in Computing Systems},
articleno = {384},
numpages = {23},
location = {New Orleans, LA, USA},
series = {CHI '22}
}

@INPROCEEDINGS{cho2017anchoring,
  author={Cho, Isaac and Wesslen, Ryan and Karduni, Alireza and Santhanam, Sashank and Shaikh, Samira and Dou, Wenwen},
  booktitle={2017 IEEE Conference on Visual Analytics Science and Technology (VAST)}, 
  title={The Anchoring Effect in Decision-Making with Visual Analytics}, 
  year={2017},
  volume={},
  number={},
  pages={116-126},
  doi={10.1109/VAST.2017.8585665}}

@misc{apnorc,
    title={{Young Adults are Leading the Way in AI Adoption - AP-NORC}},
    howpublished={\url{https://apnorc.org/projects/young-adults-leading-the-way-in-ai-adoption/}}
}

@ARTICLE{ragan2015characterizing,
  author={Ragan, Eric D. and Endert, Alex and Sanyal, Jibonananda and Chen, Jian},
  journal={IEEE Transactions on Visualization and Computer Graphics}, 
  title={Characterizing Provenance in Visualization and Data Analysis: An Organizational Framework of Provenance Types and Purposes}, 
  year={2016},
  volume={22},
  number={1},
  pages={31-40},
  doi={10.1109/TVCG.2015.2467551}}

@inproceedings{arawjo2024chainforge,
    author = {Arawjo, Ian and Swoopes, Chelse and Vaithilingam, Priyan and Wattenberg, Martin and Glassman, Elena L.},
    title = {ChainForge: A Visual Toolkit for Prompt Engineering and LLM Hypothesis Testing},
    year = {2024},
    isbn = {9798400703300},
    publisher = {Association for Computing Machinery},
    address = {New York, NY, USA},
    url = {https://doi.org/10.1145/3613904.3642016},
    doi = {10.1145/3613904.3642016},
    booktitle = {Proceedings of the 2024 CHI Conference on Human Factors in Computing Systems},
    articleno = {304},
    numpages = {18},
    location = {Honolulu, HI, USA},
    series = {CHI '24}
}

@inproceedings{wadinambiarachchi2024effects,
    author = {Wadinambiarachchi, Samangi and Kelly, Ryan M. and Pareek, Saumya and Zhou, Qiushi and Velloso, Eduardo},
    title = {The Effects of Generative AI on Design Fixation and Divergent Thinking},
    year = {2024},
    isbn = {9798400703300},
    publisher = {Association for Computing Machinery},
    address = {New York, NY, USA},
    url = {https://doi.org/10.1145/3613904.3642919},
    doi = {10.1145/3613904.3642919},
    booktitle = {Proceedings of the 2024 CHI Conference on Human Factors in Computing Systems},
    articleno = {380},
    numpages = {18},
    location = {Honolulu, HI, USA},
    series = {CHI '24}
}

@inbook{adams2015conducting,
    author = {Adams, William C.},
    publisher = {John Wiley \& Sons, Ltd},
    isbn = {9781119171386},
    title = {Conducting Semi-Structured Interviews},
    booktitle = {Handbook of Practical Program Evaluation},
    chapter = {19},
    pages = {492-505},
    doi = {https://doi.org/10.1002/9781119171386.ch19},
    url = {https://onlinelibrary.wiley.com/doi/abs/10.1002/9781119171386.ch19},
    eprint = {https://onlinelibrary.wiley.com/doi/pdf/10.1002/9781119171386.ch19},
    year = {2015}
}

@article{zhang2025exploring,
author = {Zhang, Shuning and Wang, Hui and Yi, Xin},
title = {Exploring Collaboration Patterns and Strategies in Human-AI Co-creation through the Lens of Agency: A Scoping Review of the Top-tier HCI Literature},
year = {2025},
issue_date = {November 2025},
publisher = {Association for Computing Machinery},
address = {New York, NY, USA},
volume = {9},
number = {7},
url = {https://doi.org/10.1145/3757594},
doi = {10.1145/3757594},
journal = {Proc. ACM Hum.-Comput. Interact.},
month = oct,
articleno = {CSCW413},
numpages = {43}
}

@inproceedings{peng2026designtrace,
author = {Peng, Xiaohan and Haldar, Debanjana and Mackay, Wendy E. and Koch, Janin},
title = {DesignTrace: Exploring, Iterating and Tracking Design Alternatives with GenAI},
year = {2026},
isbn = {9798400722783},
publisher = {Association for Computing Machinery},
address = {New York, NY, USA},
url = {https://doi.org/10.1145/3772318.3791036},
doi = {10.1145/3772318.3791036},
booktitle = {Proceedings of the 2026 CHI Conference on Human Factors in Computing Systems},
articleno = {434},
numpages = {22},
location = {
},
series = {CHI '26}
}

@ARTICLE{hullman2011visualization,
  author={Hullman, Jessica and Diakopoulos, Nick},
  journal={IEEE Transactions on Visualization and Computer Graphics}, 
  title={{Visualization Rhetoric: Framing Effects in Narrative Visualization}}, 
  year={2011},
  volume={17},
  number={12},
  pages={2231-2240},
  doi={10.1109/TVCG.2011.255}}

@INPROCEEDINGS{li2025reflection,
  author={Li, Haotian and Wang, Yun and Qu, Huamin},
  booktitle={2025 IEEE Visualization and Visual Analytics (VIS)}, 
  title={{Reflection on Data Storytelling Tools in the Generative AI Era from the Human-AI Collaboration Perspective}}, 
  year={2025},
  volume={},
  number={},
  pages={21-25},
  doi={10.1109/VIS60296.2025.00009}}

@article{liao2024llms,
  title={{LLMs as Research Tools: A Large Scale Survey of Researchers’ Usage and Perceptions}},
  author={Liao, Zhehui and Antoniak, Maria and Cheong, Inyoung and Cheng, Evie Yu-Yen and Lee, Ai-Heng and Lo, Kyle and Chang, Joseph Chee and Zhang, Amy X},
  journal={arXiv preprint arXiv:2411.05025},
  year={2024}
}

@inproceedings{li2024we,
    author = {Li, Haotian and Wang, Yun and Qu, Huamin},
    title = {Where Are We So Far? Understanding Data Storytelling Tools from the Perspective of Human-AI Collaboration},
    year = {2024},
    isbn = {9798400703300},
    publisher = {Association for Computing Machinery},
    address = {New York, NY, USA},
    url = {https://doi.org/10.1145/3613904.3642726},
    doi = {10.1145/3613904.3642726},
    booktitle = {Proceedings of the 2024 CHI Conference on Human Factors in Computing Systems},
    articleno = {845},
    numpages = {19},
    location = {Honolulu, HI, USA},
    series = {CHI '24}
}

@inproceedings{tang2026human,
    author = {Tang, Yuying and Zhou, Jiayi and Li, Haotian and Xie, Xing and Ma, Xiaojuan and Qu, Huamin},
    title = {How Do Human Creators Embrace Human-AI Co-Creation? A Perspective on Human Agency of Screenwriters},
    year = {2026},
    isbn = {9798400722783},
    publisher = {Association for Computing Machinery},
    address = {New York, NY, USA},
    url = {https://doi.org/10.1145/3772318.3790300},
    doi = {10.1145/3772318.3790300},
    booktitle = {Proceedings of the 2026 CHI Conference on Human Factors in Computing Systems},
    articleno = {387},
    numpages = {19},
    location = {
    },
    series = {CHI '26}
}

@ARTICLE{shi2020calliope,
  author={Shi, Danqing and Xu, Xinyue and Sun, Fuling and Shi, Yang and Cao, Nan},
  journal={IEEE Transactions on Visualization and Computer Graphics}, 
  title={Calliope: Automatic Visual Data Story Generation from a Spreadsheet}, 
  year={2021},
  volume={27},
  number={2},
  pages={453-463},
  doi={10.1109/TVCG.2020.3030403}
}

@techreport{stolper2016emerging,
    author = {Stolper, Charles D. and Lee, Bongshin and Henry Riche, Nathalie and Stasko, John},
    title = {Emerging and Recurring Data-Driven Storytelling Techniques: Analysis of a Curated Collection of Recent Stories},
    year = {2016},
    month = {April},
    howpublished = {Technical Report MSR-TR-2016-14},
    institution = {Microsoft Research},
    url = {https://www.microsoft.com/en-us/research/publication/emerging-and-recurring-data-driven-storytelling-techniques-analysis-of-a-curated-collection-of-recent-stories/},
    number = {MSR-TR-2016-14},
}

@inproceedings{sefelin2003paper,
    author = {Sefelin, Reinhard and Tscheligi, Manfred and Giller, Verena},
    title = {Paper prototyping - what is it good for? a comparison of paper- and computer-based low-fidelity prototyping},
    year = {2003},
    isbn = {1581136374},
    publisher = {Association for Computing Machinery},
    address = {New York, NY, USA},
    url = {https://doi.org/10.1145/765891.765986},
    doi = {10.1145/765891.765986},
    booktitle = {CHI '03 Extended Abstracts on Human Factors in Computing Systems},
    pages = {778–779},
    numpages = {2},
    location = {Ft. Lauderdale, Florida, USA},
    series = {CHI EA '03}
}


\end{document}